\documentclass[journal]{IEEEtran}

\usepackage{amsmath,amsfonts,bm}

\def\eqref#1{equation~\ref{#1}}

\def\1{\bm{1}}

\def\va{{\bm{a}}}

\def\vd{{\bm{d}}}
\def\ve{{\bm{e}}}
\def\vf{{\bm{f}}}

\def\vx{{\bm{x}}}

\def\mA{{\bm{A}}}

\def\mS{{\bm{S}}}

\DeclareMathAlphabet{\mathsfit}{\encodingdefault}{\sfdefault}{m}{sl}
\SetMathAlphabet{\mathsfit}{bold}{\encodingdefault}{\sfdefault}{bx}{n}
\newcommand{\tens}[1]{\bm{\mathsfit{#1}}}
\def\tA{{\tens{A}}}

\def\tL{{\tens{L}}}

\def\emS{{S}}

\usepackage{hyperref}
\usepackage{url}
\usepackage{pifont}

\usepackage{balance}

\usepackage{graphicx}
\usepackage{subcaption}
\usepackage{booktabs}
\usepackage{color, colortbl}
\usepackage{multirow}
\usepackage{wrapfig}

\usepackage[ruled,linesnumbered,vlined]{algorithm2e}
\usepackage{xspace}
\usepackage{setspace}
\usepackage[numbers]{natbib}

\usepackage{enumitem}
\setlist[itemize]{leftmargin=*}
\setlist[enumerate]{leftmargin=*}

\usepackage{tcolorbox}
\newenvironment{answerbox}{
\begin{tcolorbox}[colback=blue!5!white,colframe=blue!5!white,arc=0mm,left=1.5mm,right=1.5mm,top=0mm,bottom=0mm]
}
{
\end{tcolorbox}
}

\definecolor{mygreen}{rgb}{1, 0, 0.6}
\newcommand{\code}[1]{{\ttfamily \small #1}}

\usepackage{listings}
\usepackage{xcolor}

\definecolor{codegreen}{rgb}{0,0.6,0}
\definecolor{codegray}{rgb}{0.5,0.5,0.5}
\definecolor{codepurple}{rgb}{0.58,0,0.82}
\definecolor{backcolour}{rgb}{0.95,0.95,0.92}

\definecolor{RowGray}{gray}{0.9}

\lstdefinestyle{mystyle}{
    backgroundcolor=\color{backcolour},
    commentstyle=\color{codegreen},
    keywordstyle=\color{magenta},
    numberstyle=\tiny\color{codegray},
    stringstyle=\color{codepurple},
    basicstyle=\ttfamily\footnotesize,
    breakatwhitespace=false,
    breaklines=true,
    captionpos=b,
    keepspaces=true,
    numbers=left,
    numbersep=5pt,
    showspaces=false,
    showstringspaces=false,
    showtabs=false,
    tabsize=2
}
\usepackage{xcolor}
\definecolor{celadon}{rgb}{0.67, 0.88, 0.69}

\makeatletter
\lst@AddToHook{PreSet}{\normallineskiplimit=0pt}
\makeatother

\newcommand{\corrHumanvsModel}{+0.23}

\newcommand{\cohenIRAHumanAnswers}{0.711}
\newcommand{\cohenIRACodeGenAnswers}{0.898}
\newcommand{\cohenIRAGPTJAnswers}{0.833}
\newcommand{\cohenIRAInCoderAnswers}{0.783}

\newcommand{\nEyeTrackingParticipants}{25}
\newcommand{\nEyeTrackingSessions}{92}

\newcommand{\datasetPercCpp}{17.4\%}
\newcommand{\datasetPercCs}{43.5\%}
\newcommand{\datasetPercPy}{39.1\%}
\newcommand{\numberOfSessionsInCS}{40}
\newcommand{\numberOfSessionsInPY}{36}
\newcommand{\numberOfSessionsInCPP}{16}

\newcommand{\meanEffectiveTimeSpentPerTaskMinutes}{4.92}

\newcommand{\meanEffectiveTimeSpentPerUserMinutes}{18.54}

\newcommand{\meanNumberOfFixationsPerTask}{603.66}

\newcommand{\medianNumberOfParticipantsPerCodeSnippet}{7}

\newcommand{\medianNumberOfParticipantsPerCodeSnippetQuestion}{3}

\begin{document}

\title{Follow-up Attention: An Empirical Study of Developer and Neural Model Code Exploration}

\author{Matteo Paltenghi, Rahul Pandita, Austin Z. Henley, Albert Ziegler
\thanks{Matteo Paltenghi is with the University of Stuttgart, Stuttgart, Germany. E-mail: mattepalte@live.it. Work done while at GitHub Next for a research internship.
Rahul Pandita and Albert Ziegler are with GitHub Inc, San Francisco, CA, USA. E-mail: \{rahulpandita, wunderalbert\}@github.com.
Austin Z. Henley is with Microsoft Research, Redmond, WA, USA. E-mail: azh321@gmail.com.}
}

\maketitle

\begin{abstract}

Recent neural models of code, such as OpenAI Codex and AlphaCode, have demonstrated remarkable proficiency at code generation due to the underlying attention mechanism.
However, it often remains unclear how the models actually process code, and to what extent their reasoning and the way their attention mechanism scans the code matches the patterns of developers.
A poor understanding of the model reasoning process limits the way in which current neural models are leveraged today, so far mostly for their raw prediction.
To fill this gap, this work studies how the processed attention signal of three open large language models - CodeGen, InCoder and GPT-J - agrees with how developers look at and explore code when each answers the same sensemaking questions about code.
Furthermore, we contribute an open-source eye-tracking dataset comprising \nEyeTrackingSessions{} manually-labeled sessions from \nEyeTrackingParticipants{} developers engaged in sensemaking tasks.
We empirically evaluate five heuristics that do not use the attention and ten attention-based post-processing approaches of the attention signal of CodeGen against our ground truth of developers exploring code, including the novel concept of \textit{follow-up attention} which exhibits the highest agreement between model and human attention.
Our follow-up attention method can predict the next line a developer will look at with 47\% accuracy. This outperforms the baseline prediction accuracy of 42.3\%, which uses the session history of other developers to recommend the next line. These results demonstrate the potential of leveraging the attention signal of pre-trained models for effective code exploration.

\end{abstract}

\section{Introduction}
\label{sec:introduction}

Large language models (LLMs) pre-trained on code such as Codex~\cite{chenEvaluatingLargeLanguage2021}, CodeGen~\cite{nijkampConversationalParadigmProgram2022a}, and AlphaCode~\cite{liCompetitionLevelCodeGeneration2022} have demonstrated remarkable proficiency at program synthesis and competitive programming tasks.
Yet our understanding of why they produce a particular solution is limited.
In large-scale practical applications, the models are often used for their prediction alone, i.e., as generative models, and the way they reason about code internally largely remains untapped.

These models are often based on the attention mechanism \cite{bahdanauNeuralMachineTranslation2015}, a key component of the transformer architecture \cite{vaswaniAttentionAllYou2017}.
Besides providing substantial performance benefits, attention weights have been used to provide interpretability of neural models \cite{linStructuredSelfAttentiveSentence2017, vashishthAttentionInterpretabilityNLP2019a,paltenghiThinkingDeveloperComparing2021}.
Additionally, existing work \cite{wanWhatTheyCapture2022, vigAnalyzingStructureAttention2019, zhangWhatDoesTransformer2022, allamanisLearningRepresentPrograms2018a} also suggests that the attention mechanism reflects or encodes objective properties of the source code processed by the model.
We argue that just as software developers consider different locations in the code individually and follow meaningful connections between them, the self-attention of transformers connects and creates information flow between similar and linked code locations. This raises a question:
\begin{center}
\emph{Are human attention and model attention comparable? And if so, can the knowledge about source code conveyed by the attention weights of neural models be leveraged to support code exploration?}
\end{center}

Although there are other observable signals that might capture the concept of relevance, such as gradients-based~\cite{yuanExplainingInformationFlow2021, cheferTransformerInterpretabilityAttention2021} or layer-wise relevance propagation~\cite{montavonLayerWiseRelevancePropagation2019}, this work focuses on approaches using only the attention signal.
The reasons for this choice are two: (1) almost all state-of-the-art models of code are based on the transformer block~\cite{vaswaniAttentionAllYou2017}, and the attention mechanism is ultimately its fundamental component, so we expect the corresponding attention weights to carry directly meaningful information about the models' decision process; (2) attention weights can be extracted almost for free during the generation with little runtime overhead since the attention is computed automatically during a single forward pass.

Answering the main question of this study requires a dataset tracking developers' attention. In this work, we use visual attention as a proxy for the elements to which developers are paying mental attention while looking at code.
However, the existing datasets of visual attention are not suitable for our purposes.
Indeed, they either put the developers in an unnatural, and thus possibly biasing, environment where most of the vision is blurred~\cite{paltenghiThinkingDeveloperComparing2021}, requiring participants to move the mouse over tokens to reveal them, or they contain few and very specific code comprehension tasks~\cite{bednarikEMIPEyeMovements2020a} on code snippets too short to exhibit any interesting code navigation pattern. This blurring method can introduce bias by forcing unnatural interactions, potentially affecting how developers naturally explore and understand code.
To address these limitations and stimulate developers to not only glance at code, but also to deeply reason about it, we prepare an ad-hoc code understanding assignment called the \textit{sensemaking task}.
This involves questions on code, including mental code execution, side-effects detection, algorithmic complexity, and deadlock detection.
Moreover, using eye-tracking, we collect and share a dataset of \nEyeTrackingSessions{} valid sessions with developers.

On the neural model side, motivated by some recent successful applications of few-shot learning in code generation and code summarization~\cite{bareissCodeGenerationTools2022, ahmedFewshotTrainingLLMs2023} and even zero-shot in program repair~\cite{xiaLessTrainingMore2022}, the sensemaking task is designed to be a zero-shot task for the model with a specific prompt that triggers it to reason about the question at hand.
Then we query three LLMs of code, namely CodeGen~\cite{nijkampConversationalParadigmProgram2022a}, InCoder~\cite{friedInCoderGenerativeModel2022} and GPT-J~\cite{gpt-j} on the same sensemaking task and compare their \textit{attention signal}\footnote{Attention signal refers to the attention weights produced during a forward pass by the transformer blocks.} to the attention of developers.
The correlation with CodeGen, the largest model, is the highest among the LLMs studied (r=\corrHumanvsModel{}), motivating the use of raw and processed versions of CodeGen's attention signal for code exploration.
To that end, we experimentally evaluate how well existing and novel attention post-processing methods align with the code exploration patterns derived from our dataset's chronological sequence of eye-fixation events.
To the best of our knowledge, this work is the first to investigate the attention signal of these pre-trained models to support code exploration, a specific code-related task, directly related to code reading work~\cite{blascheckVisuallyAnalyzingEye2019, busjahnEyeMovementsCode2015}.

We empirically demonstrate that post-processing methods based on the attention signal can be well aligned with the way developers explore code.
In particular, using the novel concept of \textit{follow-up attention}, we achieve the highest overlap with the developers' ground truth on which line to explore next.

\noindent\textbf{Contributions}: This paper makes the following contributions:
\begin{itemize}[noitemsep, leftmargin=*, topsep=0pt]
   \item[$\star$] \textbf{Sensemaking Task} A novel task and setup to deepen our understanding of how the LLM attention connects to the temporal sequence of location shifts regarding developer focus.
   \item[$\star$] \textbf{Eye-Tracking Dataset} A novel dataset of \nEyeTrackingSessions{} eye tracking sessions of \nEyeTrackingParticipants{} developers engaged in sensemaking tasks while using a common code editor with code written in three popular programming languages (Python, C++, and C\#).
   \item[$\star$] \textbf{Follow-up Attention} The analytical formula for \textit{follow-up attention}, a novel post-processing approach derived solely from the attention signal, which aligns well with the developer interaction of which line to look at next when exploring code.
   \item[$\star$] \textbf{Empirical Study} The first comparison of both effectiveness and visual attention of LLMs and developers when reasoning on sensemaking questions.
   An empirical evaluation comprising ten post-processing approaches of the attention signal, five heuristics, and an ablation study of the follow-up attention against the collected ground truth of developers exploring code.
\end{itemize}

\section{Related Work}
\label{sec:related_work}
This section provides an overview of related work around the explanatory role of attention and previous studies of the attention of neural models and developers when reasoning on code.

\textbf{Attention as explanation.}
Initially, preliminary work~\cite{jainAttentionNotExplanation2019} studying attention weights of recurrent neural models has found that the attention weights do not always agree with other explanation methods and that alternative weights can be adversarially constructed while still preserving the same model prediction.
However, in response, Wiegreffe and Pinter~\cite{wiegreffeAttentionNotNot2019} have shown how the alternative attention weights can be constructed only per a single instance prediction, whereas obtaining a model which is consistently wrong in its explanations is very unlikely to happen.
On the same line, Tutek and Šnajder~\cite{tutekStayingTrueYour2020} have proposed four regularization methods to mitigate the adversarial exploitation of attention weights for recurrent models, including the use of residual connections which are natively embedded into transformers~\cite{vaswaniAttentionAllYou2017}, the building blocks of the LLMs studied in this work.
To further corroborate this connection between attention and explanation, Rabin et al.~\cite{rabinUnderstandingNeuralCode2021a} have shown how even Sivand, an explainability technique based on program simplification, pinpoint important tokens which largely overlap with those reported by the attention mechanism.

\textbf{Attention studies of neural models of code.}
Paltenghi and Pradel~\cite{paltenghiThinkingDeveloperComparing2021} have compared the attention weights of neural models of code and developers' visual attention when performing a code summarization task, and found a strong positive correlation on the \textit{copy attention} mechanism for an instance of a pointer network~\cite{vinyalsPointerNetworks2015}.
Further works \cite{wanWhatTheyCapture2022, zhangWhatDoesTransformer2022} have then shown how the attention weights of pre-trained models on source code capture important properties of the abstract syntax tree of the program.
However, none of them considered the use of the attention signal for a code-related task, such as code exploration.
Moreover, they are limited to relatively small self-attention transformer models, whereas we study the attention of CodeGen~\cite{nijkampConversationalParadigmProgram2022a}, InCoder~\cite{friedInCoderGenerativeModel2022} and GPT-J~\cite{gpt-j}, large generative models with masked self-attention.

\textbf{Eye-Tracking Studies}
Turner et al.~\cite{turnerEyetrackingStudyAssessing2014} conducted an eye-tracking study involving 38 students fixing or describing five simple Python and C++ programs (5-13 LoC) showing that the fixation duration is comparable between the two languages.
Beelders~\cite{beeldersEyetrackingAnalysisSource2022} has qualitatively observed the eye movement of 36 students and four lecturers when reading and mentally executing a short C\# program (12 LoC).
An eye-tracking dataset with 216 participants has been collected by \cite{bednarikEMIPEyeMovements2020a}, however, they only consider two short snippets (11-22 LoC) of code, since they do not support scrolling.
Similarly, Blascheck and Sharif~\cite{blascheckVisuallyAnalyzingEye2019} and Busjahn et al.~\cite{busjahnEyeMovementsCode2015} have studied the reading order in C++ and Java code comprehension task focusing on six small programs that could fit into a single screen, whereas we consider longer snippets and a much larger dataset of 45 unique tasks.
Sharifi et al.~\cite{sharafiEyesCodeStudy2022} have recently studied code navigation strategies on Java code with eye tracking involving 36 participants focusing on the bug fixing process, however, we study the sensemaking task which might elicit a different kind of reasoning compared to bug-fixing.
To more closely mimic real-world setups in integrated development environments (IDEs), Guarnera et al.~\cite{guarneraITraceEyeTracking2018} propose iTrace, an eye-tracking plugin for IDEs that can track developers' eye movements in more realistic and dynamic coding environments beyond a single screen of code.
Further studies, including Fakhoury et al.~\cite{fakhouryGazelSupportingSource2021a}, have proposed Gazel, an IDE plugin that supports eye tracking in the context of source code editing.
Following this latest trend, we also use an IDE plugin to collect the eye-tracking data, allowing for a more realistic coding environment.

\section{Sensemaking Task}
\label{sec:sense_making_task}

\begin{figure}[t]
   \begin{lstlisting}[language=Python, numbers=left]
# ************************************************
# The following code reasons about triangles in the geometrical sense.
class point:
      def __init__(self, x, y):
         self.x = x
         self.y = y
def square(x):
      return x * x
def order(a, b, c):
      copy = [a, b, c]
      copy.sort()
      return copy[0], copy[1], copy[2]
...
p1 = point(0, 0)
p2 = point(1, 1)
p3 = point(1, 2)
classifyTriangle(p1, p2, p3)
# Question: What could happen if the call to `order()` were omitted from
# `classifyTriangle`?
# Answer:
\end{lstlisting}
\caption{Example of sensemaking task with code and question to be answered in the bottom comment. Completely empty lines have been removed for space reasons.}
\label{fig:sense_making_task}
\end{figure}

To study developers' and models' attention, we prepare a code understanding task called \textit{sensemaking task} because the developer has to ``make sense'' of code to answer the question correctly.
One sensemaking task is contained in a single source code file $p$ composed of four sections: (1) a brief description of the context of the main code snippet (e.g., \textit{The following code reasons about triangles in the geometrical sense.}), (2) the main code snippet, either sourced on the internet or written from scratch by the authors, (3) a sensemaking question to stimulate the reasoning (i.e., \code{Question:}), and (4) a final prompt to trigger the model's answer (i.e., \code{Answer:}).
Note that all the sections except the main snippet are in the form of code comments.
Figure~\ref{fig:sense_making_task} shows an example task, whereas the full list of questions can be seen in the Table~\ref{tab:sense_making_questions}.

To source the tasks for our study, we rely on \emph{GeeksforGeeks}\footnote{\url{https://www.geeksforgeeks.org/}}, a well-known website for programming education and practice. This website offers a variety of problem statements that are commonly used in typical technical interviews by modern software companies, as shown by previous research~\cite{behroozi2019hiring}. Therefore, we expect that the software developers would have some familiarity with the type of these programs. We then create specific sense-making questions about these programs, inspired by the kind of questions that an interviewer might pose, such as asking about the output, complexity, correctness, or code modification.
Indeed, many of our questions are concrete instances of question templates such as ``What is the purpose of the code?'' (\code{nqueens\_Q1}), ``What is the program supposed to do?'' (\code{tree\_Q3}) or ``What code could have caused this behavior?'' (\code{triangle\_Q1}), which also have been identified as questions that software engineers often ask themselves in a real working setting~\cite{koInformationNeedsCollocated2007}.
To stimulate code exploration, many of them are also instances of reachability questions~\cite{latozaDevelopersAskReachability2010}; namely, they involve the search over all feasible paths of a program to locate target statements matching search criteria.
Some examples of these are ``What are the implications of this change?'' (\code{triangle\_Q3}) or ``How does application behavior vary in these different situations that might occur?'' (\code{triangle\_Q2}, \code{tree\_Q1}, \code{multithread\_Q3}).
We prepare five main snippets and create three unique questions for each of them. Then we translate the same task into three programming languages: Python, C++, and C\#.
In total, we have 45 unique tasks.
Although the sensemaking task includes questions that might have also been asked in studies focused on code comprehension~\cite{wyrich40YearsDesigning2023}, the main difference is those studies typically restrict the scope of their questions to either bottom-up~\cite{wyrich40YearsDesigning2023} or top-down comprehension~\cite{shaftRelevanceApplicationDomain1995} tasks.
Whereas, in our sensemaking task, beside code snippet and question, participants receive also the header of the file with some contextual information, which creates an unusual blend of bottom-up and top-down comprehension tasks which is typically not seen in code comprehension studies which focus on either one or the other.
This decision is motivated by our goal of stimulating code exploration, where the participants have to integrate different pieces of information at different locations and create an integrated mental model.

\begin{table*}[h]
   \small
   \centering
   \begin{tabular}{p{0.13\textwidth}p{0.1\textwidth}p{0.06\textwidth}p{0.63\textwidth}}\toprule
      Snippet Name &                     Content &     LoC &                                                                                                                                                           Question \\

   \midrule
   nqueens\_Q1 &            \multirow{3}{0.1\textwidth}{N queens problem} &  78-100 &                                                                                                                                   What does `solveNQ(-13)` return? \\
   nqueens\_Q2 &            &  78-101 &                                                                                                        What are valid dimensions and values for the array `board`? \\
   nqueens\_Q3 &            &  78-100 &                                                                                               How would you expect the run time of `solveNQ(n)` to scale with `n`? \\
   \midrule
         hannoi\_Q1 &     \multirow{3}{0.1\textwidth}{Tower of Hanoi problem} &   28-49 &                                                                                        How does the algorithm moves disks from the starting rod to the ending rod? \\
         hannoi\_Q2 &       &   26-47 &                                                                                                                           Which is the base case of the algorithm? \\
         hannoi\_Q3 &       &   28-50 &                                                                         Which is the name of the auxiliary rod in the call TowerOfHanoi(n, 'Mark', 'Mat', 'Luke')? \\
   \midrule
    multithread\_Q1 &   \multirow{3}{0.1\textwidth}{Consumer-producer threads} & 106-116 & Is it possible that consumer and producers threads end up in a deadlock state; namely they both wait for each other to finish, but none of them is doing anything? \\
    multithread\_Q2 &   & 104-112 &                                                  Is there any line of code in the consumer or producer code that will never be executed? If yes, report it below. \\
    multithread\_Q3 &   & 104-113 &                            Will the queue object ever raise an exception in this program? If yes, which condition(s) should be met for the exception to be raised? \\
   \midrule
           tree\_Q1 & \multirow{3}{0.1\textwidth}{Recursive tree construction} &   87-99 &                                                                         How many calls to `constructTreeUtil` will `constructTree([1, 2, 3], [ 1, 2, 3], 2)` make? \\
           tree\_Q2 &  &   87-99 &                                                                                Under which conditions could the check `if i <= h` in `constructTreeUtil` be false? \\
           tree\_Q3 &  &  89-101 &              A part of the code you don't have direct access to has called `constructTree` with unknown parameters. What can you find out about  those parameters? \\
   \midrule
       triangle\_Q1 &     \multirow{3}{0.1\textwidth}{Triangle classification} &  66-112 &                                              Which of the functions have side effects (namely it modifies some state variable value outside its local environment? \\
       triangle\_Q2 &      &  66-113 &                                                                               Which output will you get for the three points [1, 2], [1, 3], and [1, 4]? \\
       triangle\_Q3 &     &  66-112 &                                                                                   What could happen if the call to `order()` were omitted from `classifyTriangle`? \\
    \bottomrule
    \end{tabular}
   \caption{Code snippets and related questions for each sensemaking task.}
   \label{tab:sense_making_questions}
   \end{table*}

\textbf{Neural Model's Task}. We feed the entire source file of a single task as input, also referred to as \emph{prompt}, to the generative model and query it for three different answers in the form of text completion.
A model processes the input file $p$ by splitting it in tokens via a deterministic tokenizer ($p={t_1, ..., t_n}$) and then generates a sequence of tokens as output, as shown on the left of Figure~\ref{fig:overview}.
We allow the models to generate an answer of length 100 tokens at maximum, which is more than enough to respond to all the questions.
We use three widely used open source pre-trained models namely: CodeGen~\cite{nijkampConversationalParadigmProgram2022a} in its language-agnostic variant\footnote{\code{CodeGen-16-multi} from \url{https://github.com/salesforce/codegen}}, InCoder~\cite{friedInCoderGenerativeModel2022} and GPT-J~\cite{gpt-j}, all in their largest variants of 16B, 6B and 6B of parameters respectively.
To query the model multiple times we use the temperature sampling strategy with a temperature of 0.2.

\textbf{Developers' Task}.
We recruit \nEyeTrackingParticipants{} software developers via direct contacts at a large software company, ranging from interns to more senior software engineers, thus having diverse degrees of familiarity with software development and programming.
We track the eye gaze of each participant during a 19 minutes session (on average) while they answer as many questions as possible, typically three or four.
We ensure they see each main code snippet only once to avoid bias in answering a question on a snippet they have already explored in a previous task.
The eye-tracking setup is calibrated at the beginning of each task to ensure consistent data collection.

\section{Problem Formulation}
\label{sec:problem_formulation}

The majority of modern large language models (LLMs) are based on the architecture of generative pre-trained transformers (GPT)~\cite{radfordImprovingLanguageUnderstanding}, such as Codex~\cite{chenEvaluatingLargeLanguage2021}, CodeGen~\cite{nijkampConversationalParadigmProgram2022a}, and AlphaCode~\cite{liCompetitionLevelCodeGeneration2022}.
Self-attention is a mechanism used in these models that allows each processed token to weigh its own importance with respect to other tokens in the same sequence, enabling the model to capture relationships and dependencies within the sequence.
In particular, the representation of each token can incorporate information from tokens that come earlier in the sequence, and on the contrary cannot incorporate information from tokens that come later in the sequence.
In this work, when a token $A$ incorporates information from another token $B$, we say that $A$ \textit{attends} to $B$, or equally that token $A$ pays \textit{attention} to token $B$.
This attention is usually quantified by a scalar value, called \textit{attention weight}, which is computed by the model in its attention mechanism.

\begin{figure*}[t]
   \centering
   \includegraphics[width=0.92\textwidth]{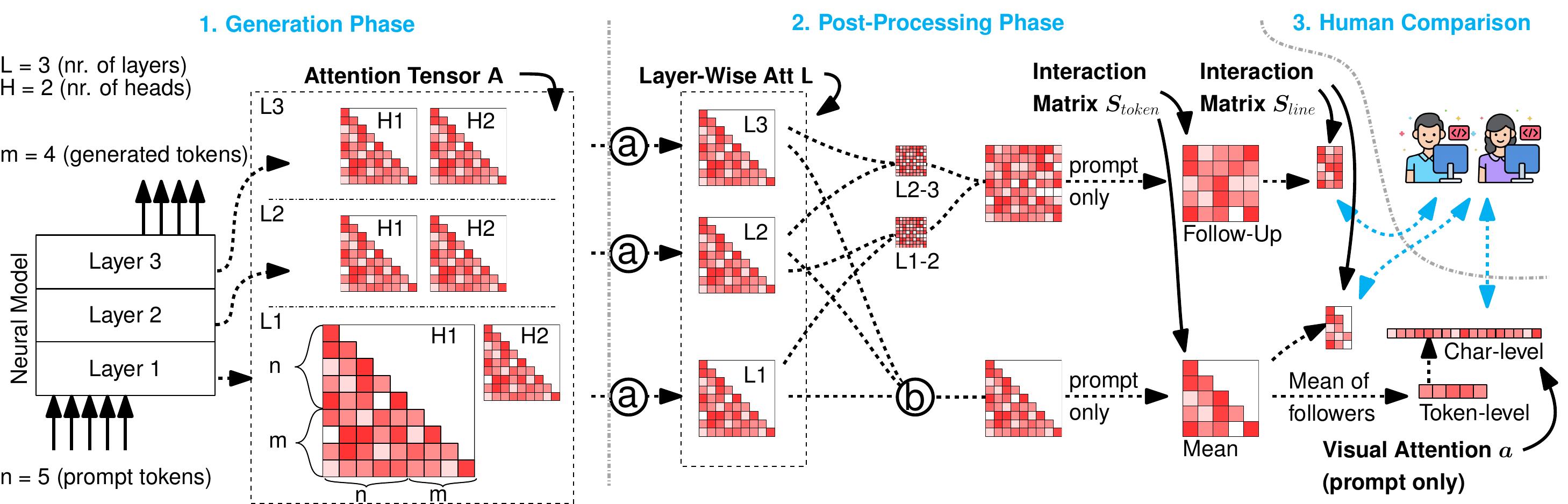}
   \caption{Overview of the three extraction functions for the visual attention vector and the interaction matrix, both follow-up and mean. Note that $a$ and $b$ represent specific aggregation functions as explained in the text (e.g., mean, max or sum). The darker the red color, the more attention is paid to by token on the row $i$ to the token on the column $j$.}
   \label{fig:overview}
 \end{figure*}

When the model takes as input a sequence of $x$ tokens, the attention mechanism is applied to each token in the sequence.
Figure~\ref{fig:overview} on the left shows a toy example with a model of three layers and two attention heads, together with the attention generated by the model.
For each token, the attention is computed sequentially through the $L$ \textit{layers} of the neural model and, at each layer, the attention is computed in parallel $H$ times, once for each sub-network called \textit{attention head}.
Fixing a combination of layer and head, the attention given by $i$-th token to the other tokens of the sequence can be represented by a vector of weights: $\displaystyle \va_i = (a_{i,1}, a_{i,2}, ... , a_{i,i}, 0_{i,i+1}, ..., 0_{i, x})$ where $a_{i,j}$ is the weight given by token at position $i$ to token at position $j$.
Note that the token cannot attend any token that come later in the sequence, thus the weights $a_{i,j}$ are zero for $j>i$.
Stacking the attention vectors one after the other as row, we obtain an \textit{attention matrix} $\displaystyle \mA = (\va_1, \va_2, ..., \va_x)$ for the specific combination of layer and attention head, note that it is a lower triangular matrix.

Thus, when the input file comprising $n$ tokens ($t_1, ..., t_n$) is fed to the model $f$, beside a predicted answer of $m$ newly generated tokens ($t_{n+1}, ..., t_{n+m}$), the model also computes an attention tensor $\tA$ of shape $(L, H, n+m, n+m)$, where $L$ is the number of layers and $H$ is the number of attention heads.
In particular, when comparing developers' and the model's attention, we focus on studying the attention weights referring to the prompt tokens only, even if some post-processing approach may use the entire tensor $\tA$.

Note that by construction, not all tokens can attend to all other tokens, thus we define the notions of \emph{followers} of a token $t_i$ as the set of tokens that can pay attention to $t_i$.
This set is defined as $\displaystyle \mathcal{F}(t_i) = \{t_j \mid j > i\}$, where the subscript represents the position of the token in the sequence.

\subsection{Views of Attention}
In our problem formulation, we model an extraction function $g$ that takes as input the attention tensor $\tA$ and returns either a measure of \textit{how much attention the model pays to each part of the prompt} or a measure of \textit{how much each part is linked to other parts} of the prompt.
Depending on the case, we refer to the outputs as \textit{visual attention vector} or \textit{interaction matrix} respectively.

\textbf{Visual Attention Vector.}
It is a static view telling us which part of the input is important for the model when solving the sensemaking task.
We define a \textit{visual attention} of a model as a vector $\displaystyle \va = (a_1, ..., a_{c})$ over the $c$ characters of the prompt, where each $a_i$ intuitively tells us how much attention was given to that the $i$-th character when solving the task.
We use $g_{viz}(\tA)$ to model a function that takes as input the attention tensor $\tA$ and returns a visual attention vector $\va$.

\textbf{Interaction Matrix.}
It is a dynamic view that tells us, given a position in the prompt, which other position of the prompt is more deeply connected to it.
We define an interaction matrix $\displaystyle \mS$ as a right stochastic matrix with size $n \times p$ where $n$ is the number of tokens in the prompt and $p$ is the number of admissible target positions in the prompt.
We distinguish two kinds of interaction matrices depending on the granularity
of the target position $p$, either pointing to another token or line in the source code (the latter being of interest primarily with developer tooling in mind, which is often line based).
Respectively, we call them: (1) \textit{token-level}, where $\displaystyle \mS$ has size $n \times n$ where $n$ is the number of tokens in the prompt; (2) \textit{line-level}, where $\displaystyle \mS$ has size $n \times n_l$ where $n_l$ is the number of lines in the prompt.
We use $g_{token}(\tA)$ and $g_{line}(\tA)$ to model two functions that take as input the tensor $\tA$ and output an interaction matrix, either $\mS_{token}$ or $\mS_{line}$ respectively.

\section{Extraction Functions}
\label{sec:extraction_functions}
We investigate two algorithms for extracting the visual attention vector and four for the interaction matrix.

\subsection{Attention Extraction Overview}
Figure~\ref{fig:overview} illustrates the process of querying the model and extracting its attention signals, leading to the comparison with human developers.
In particular, it shows how to the attention tensor $\tA$ is derived by querying the model, and how to post process it to extract both the interaction matrix $\mS_{token}$ and the visual attention vector $\va$.
It is split in three phases: generation, post-processing, and human comparison.
In the generation phase, a neural model with $L$ layers and $H$ attention heads processes $n$ prompt tokens to generate both $m$ tokens and an attention tensor $\tA$. Here, the model has three layers and two attention heads, handling five prompt tokens and generating four new tokens.
During the post-processing phase, the attention tensors are aggregated to form interaction matrices $\mS_{token}$ using two techniques: mean aggregation and follow-up attention, as explained in Section~\ref{sec:extract_functions_interaction_matrix}.
Note that the matrices are cut to consider only the attention to the $n$ tokens of the prompt.
Then, the interaction matrix from token-to-token is converted to the line-level interaction matrix by aggregating the attention weights of the tokens belonging to the same line, obtaining $\mS_{line}$.
At this point, the visual attention vector $\va$ is extracted from the $\mS_{token}$ matrix (see Section~\ref{sec:extract_functions_visual_attention}) and converted to the character level.
In the last phase, the interaction matrices $\mS_{line}$ and the visual attention vectors $\va$ are compared with human data collected via eye-tracking.

\subsection{Visual Attention Vector Extraction}
\label{sec:extract_functions_visual_attention}

We introduce two alternative approaches called \textit{attention mean} and \textit{attention max} to condense the attention tensor $\tA$ to the visual attention vector $\displaystyle \va$, namely to implement $g_{viz}(\tA): \tA \rightarrow \va$. The first approach is visualized in the bottom part of Figure~\ref{fig:overview}.

\textbf{Attention mean.}
It aggregates over all the layers $L$ and attention heads $H$ by taking the average attention weight for each token position.
After keeping only the prompt tokens, this step outputs a matrix $\mA$ with shape $(n, n)$ where each element $\mA_{i,j}$ is the average attention paid by the $i$-th token to the $j$-th token in across all layers and heads.
Note that it is a lower triangular matrix because a token cannot attend tokens coming after it by construction.
Then, we compute the mean of each column excluding the zeros to avoid penalizing more recent tokens with fewer followers.
This step corresponds to represent each token $t_i$ with the average attention given to it by its followers $\mathcal{F}(t_i)$, thus we call the step \textit{mean of followers}.
It outputs a token-level visual attention $\va$ vector that is converted to character-level vector, by dividing the attention weight on a single token in \textit{equal shares} among all its characters.

\textbf{Attention max.}
This approach differs from the previous one in how it condenses layers and heads in the first step, replacing the mean with the max function to favor the extremely positive signals appearing only in one or few layers and heads; the rest is unchanged.

\subsection{Interaction Matrix Extraction}
\label{sec:extract_functions_interaction_matrix}

We study four approaches: \textit{mean}, \textit{max}, \textit{rollout} and \textit{follow-up attention}.
Apart from the rollout attention, which has been introduced by \cite{abnarQuantifyingAttentionFlow2020}, the other three are either inspired by the work of \cite{paltenghiThinkingDeveloperComparing2021} or a novel contribution of this work, such as the follow-up attention.

\begin{algorithm}
\setstretch{0.6}
   \small
\caption{Follow-up Attention}
\label{alg:followup_attention_model}

\SetKwInOut{Input}{Input}
\SetKwInOut{Output}{Output}
\SetKwFunction{FMain}{FollowUpAttention}
\SetKwProg{Fn}{Function}{:}{}

\Fn{\FMain}{
   \Input{$\tA$ \tcp*[r]{Dimensions: $(L, H, n+m, n+m)$}}
   \Output{$\mS$ \tcp*[r]{Dimensions: $(n, n)$}}

   $CP \gets \emptyset$  \tcp*[r]{$CP$: Consecutive layer pairs}
   $\tL \gets \sum_{h=1}^H \tA_{h}$ \\ \label{alg:layer_wise_aggregation}
   \For{$z \gets 1$ to $L-1$}{  \label{alg:iterate_layer_pairs}
      $\mS^{(z)} \gets \emptyset$ \\
      \For{$i \gets 1$ to $n$}{  \label{alg:iterate_over_i}
         \For{$j \gets 1$ to $n$}{   \label{alg:iterate_over_j}
            $s \gets n+m-\max(i, j)$ \\
            $\vf_{i}^{(z)} \gets \tL_{z, s:, i}$ \\ \label{alg:follower_score}
            $\vf_{j}^{(z+1)} \gets \tL_{z+1, s:, j}$ \\
            $\mS_{i,j}^{(z)} \gets \frac{\vf_{i}^{(z)} \cdot \vf_{j}^{(z+1)}}{||\vf_{i}^{(z)}|| \cdot ||\vf_{j}^{(z+1)}||}$ \\ \label{alg:cosine_similarity}
         }
      }
      $CP \gets CP \cup \mS^{(z)}$ \\
   }
   $\mS \gets \sum_{\mS^{(z)} \in CP} \mS^{(z)}$ \\  \label{alg:condense_layers}

   \Return{$\mS$} \\
}
\end{algorithm}

\textbf{Attention mean.} It computes the mean among all the $L$ layers and $H$ attention heads: $g_{mean} = \frac{1}{l \cdot h} \sum_{l=1}^L \sum_{h=1}^H \tA_{l,h}$.

\textbf{Attention max.} It computes the max among all the $L$ layers and $H$ attention heads: $g_{max} =\max_{l=1}^L \max_{h=1}^H \tA_{l,h}$.

\textbf{Rollout attention.} It propagates the information contained in the attention weights layer by layer from input to output by multiplying the attention weights along multiple paths starting and ending in the same input-output pair.
Since it does not model the attention head dimension, we condense that dimension via a simple sum.
There is no a priori criterion for which attention layer should be used in the end. Thus, we sum the rollout values of all the layers.
\cite{abnarQuantifyingAttentionFlow2020} for more details.

\textbf{Follow-up attention.} It is our novel approach that centers on modeling the flow of information between subsequent layers. The intuition behind it is that follow-up attention tracks whether a token being attended to in one layer will cause a different token being attended to in the next layer. This is the model analogue of tracking humans jumping from attending one token to attending a different token next.

Algorithm~\ref{alg:followup_attention_model} summarizes the entire procedure\footnote{An optimized vectorized version is available here \url{https://github.com/githubnext/followup-attention}}, which is also represented in Figure~\ref{fig:overview}.
Similarly to the rollout attention, we aggregate the attention weights over the $H$ attention heads by summing the weights of $\displaystyle \tA$ along the attention head dimension and obtaining a \textit{layer-wise attention} $\tL$, a 3-dimensional tensor (Line~\ref{alg:layer_wise_aggregation}).
The follow-up attention explicitly models the temporal relationship between the attention weights computed at different layers since the attention weights in the later layers depend on the earlier ones.
The intuition is that the layer-after-layer transformation reflects how the models explore code through time, similar to multiple successive fixations of a developer when navigating and exploring source code.
Instead of looking at how the token gives attention to other tokens in the same layer, the follow-up attention adopts a differential approach which compares the attention \textit{received} by token $i$ at layer $z$ with the attention \textit{received} by token $j$ at layer $z-1$.
To represent this received attention, we define the \textit{follower score} $\displaystyle \vf_{i}^{(z)}$ of token $i$ at layer $z$, as the vector of the attention quota that each other token (which we call \emph{observers}) gives to token $i$ at the same layer (Line~\ref{alg:follower_score}).
Note that, similarly to the attention vector, the follower score is also a vector of real numbers and it has the same length corresponding to the input sequence length, thus representing a complementary viewpoint.
To realize the agreement between follower scores at two consecutive layers, we use the cosine similarity as a soft version of the intersection between the set of followers of the two tokens (Line~\ref{alg:cosine_similarity}).
Then we compute the follow-up attention for each ordered pair of tokens $i \rightarrow j$ (Lines~\ref{alg:iterate_over_i}-\ref{alg:iterate_over_j}) and for each pair of consecutive layers (Line~\ref{alg:iterate_layer_pairs}) and condense all layer pairs into a single matrix via sum (Line~\ref{alg:condense_layers}).
We aggregate attention over multiple layers since \cite{brunnerIdentifiabilityTransformers2020} have empirically shown how token identifiability is retained over layers, thus a generic embedding at position $\displaystyle \ve_i$ in any layer $l$ is traceable to the input embedding $\displaystyle \vx_i$ in the input sequence.

\section{Code Exploration Dataset}
\label{sec:dataset}

\begin{table}[h]
   \centering
   \caption{Participants' Professional Software Development Experience}
   \label{tab:participants_experience}
   \begin{tabular}{cccccc}
   \toprule
   \textbf{Experience (years)} & $\leq$ 1 & 1-2 & 2-4 & 4-6 & $\geq$ 10 \\
   \midrule
   \textbf{Number of Participants} & 8 & 7 & 3 & 3 & 3 \\
   \bottomrule
   \end{tabular}
\end{table}

We conduct an eye-tracking study and collect a novel dataset comprising \nEyeTrackingParticipants{} participants across \nEyeTrackingSessions{} valid sessions.
Participants’ professional software development experience is summarized in Table~\ref{tab:participants_experience}.
Additionally, 24\% of participants have more than 4 years of experience.
The dataset contains \datasetPercCpp{} (\numberOfSessionsInCPP{}) of sessions on C++ code, \datasetPercCs{} (\numberOfSessionsInCS{}) on C\# code, and \datasetPercPy{} on Python code (\numberOfSessionsInPY{}).

The sessions are single-purpose and live-monitored by an experimenter to ensure correct setup and focus on the code exploration task. While each participant has 45 minutes to solve as many tasks as possible, due to calibration and transition times, on average, they spend an average of \meanEffectiveTimeSpentPerUserMinutes{} minutes exploring code using the IDE, with an average of \meanEffectiveTimeSpentPerTaskMinutes{} minutes per single question. A pair of code snippet and question is looked at by a median number of \medianNumberOfParticipantsPerCodeSnippetQuestion{} different participants, and each code snippet is looked at by a median number of \medianNumberOfParticipantsPerCodeSnippet{} different participants. No participant is presented with the same code snippet more than once.

Each session consists of a sequence of eye fixation events $evt_{eye}$, each represented as a tuple $(t, x, y, d)$ where $t$ is the timestamp in milliseconds, $x$ and $y$ are the coordinates of the fixation point in pixels and $d$ is the duration of the fixation in milliseconds.
The average of fixations per session is \meanNumberOfFixationsPerTask{}.
Each session is recorded in Visual Studio Code\footnote{\url{https://code.visualstudio.com/}} to have a natural coding environment.
Based on the size of the parafoveal region~\cite{schotterParafovealProcessingReading2012}, each eye fixation event is converted to column and line coordinates: $evt_{eye}^{(char)} = (t, c, l, d)$ where $c$  is the column, and $l$ is the line of the original source file.

\subsection{Eye Tracking Setup}
\label{sec:setup}
To collect the eye tracking data, we use an eye tracker from GazePoint (model GP3, with 0.5 – 1 degree of visual angle accuracy), which is placed below the monitor thus not requiring the user to wear any additional device.
Note that our setup is as close as possible to a normal coding session without any invasive or unnatural methods.
The participants can see between 21 and 26 lines of code.
The screen size is 52.7 mm x 29.6 mm with a resolution of 1920x1080 pixels.
The participant seats at a fixed distance of approximately 30 cm from the screen.
The fixation are computed by the internal fixation filter of the eye tracker, which uses a custom algorithm based on displacement~\cite{VirtualEnvironmentsAdvanced1995}, using the \code{FPOG} (Fixation Point of Gaze) data stream\footnote{Manual: \url{https://www.gazept.com/dl/Gazepoint_API_v2.0.pdf}}
Eye-tracking data are pre-processed using custom code in Python (version: 3.8) described next below and openly shared (see Data Availability).

Besides collecting the eye tracking data $evt_{eye}$, our setup also collects $evt_{txt}$ coming from a custom VSCode plugin that logs the visible text on the screen.
A visible text event $evt_{txt}$ corresponds to a tuple $(t, txt, f, l )$ where $t$ is the timestamp in milliseconds, $txt$ is the visible text, $f$ is the file name shown in the code area, $l$ is the line number of the first visible line with respect to the given file.
Note that this event is crucial since we study long code snippets and allow also screen scrolling.
To ensure that we have a consistent grid mapping between pixel positions and char positions in the text, we use a monospace font, prevent partial scrolling and prevent any resizing of the code area during the experiment.
Then, for each timeframe, we map the pixel of each eye fixation event to a specific character position in the code area by using a grid over the character positions identified by a line and column.

To derive the developer attention maps from the eye tracking data, we first synchronize data from the VSCode plugin and the eye tracker, via their timestamps.
Then, we convert the fixation point of gaze $x$ and $y$ coordinates of each $\mathit{evt}_{\mathit{eye}}$ to the corresponding character line and column coordinates in the relative coordinate system of the code area.
And knowing the line number $l$, we can attribute the fixation to a specific character position in the original file.
In this way, we convert each $\mathit{evt}_{\mathit{eye}}$ to its equivalent event in character coordinates $evt_{\mathit{eye}}^{(\mathit{char})} = (t, c, l, d)$ where $t$ is the timestamp, $c$ and $l$ are the column and line coordinates with respect to the original file of the fixation point and $d$ is the duration of the fixation.

Since it is hard to tell whether, during a single fixation event, a participant is looking at a specific character or a group of neighboring characters, we attribute the developer's attention to neighboring characters.
In particular, if the developer looks at position $(c, l)$ in the original file, we augment our data by introducing new events which point to all the neighboring characters within a vertical offset $v_{\mathit{off}}$ and a horizontal offset $h_{\mathit{off}}$ from our coordinate $(c, l)$.
As a result, we replace each basic $evt_{eye}^{(char)} = (t, c, l, d)$ with the set of derived events $(t, c_{\mathit{new}}, l_{\mathit{new}}, d)$ where $c - h_{\mathit{off}} \leq c_{\mathit{new}} \leq c + h_{\mathit{off}}$ and $l - v_{\mathit{off}} \leq l_{\mathit{new}} \leq l + v_{\mathit{off}}$.
This is strictly connected to the concept of fovea region.
Indeed, as reported by \cite{schotterParafovealProcessingReading2012}, our fovea region, which is responsible for a sharp central vision, accounts for 2\textdegree~of the visual field, whereas the parafoveal region, which is used for visual search and scene perception, accounts for 5\textdegree~of the visual field.
Thus considering 5\textdegree~visual region and our screen size (527mm x 296mm), the developer can see 7.16 characters horizontally and 2.92 characters vertically.
Rounding those quantities we set $v_{\mathit{off}} = 1$ and $h_{\mathit{off}} = 4$.
This approach also contributes to mitigating any small $x$ and $y$ errors in the eye-tracking data collection.

\subsection{Ground Truth Visual Attention}
Here we borrow from the concept of \textit{human attention} proposed by \cite{paltenghiThinkingDeveloperComparing2021} and define the analogous \textit{developer attention} as the total time that a specific char was visible to the participant (i.e., in their field of vision): $\displaystyle \vd = (d_1, ..., d_{c})$ where $c$ is the number of characters in the prompt and $d_i$ is the total time that the $i$-th character was visible to the participant according to the eye tracking data.
In contrast to \cite{paltenghiThinkingDeveloperComparing2021}, we consider the char-level instead of the token level because it is more natural for our eye-tracking data.

\begin{figure}[t]
  \centering
  \includegraphics[width=0.45\textwidth]{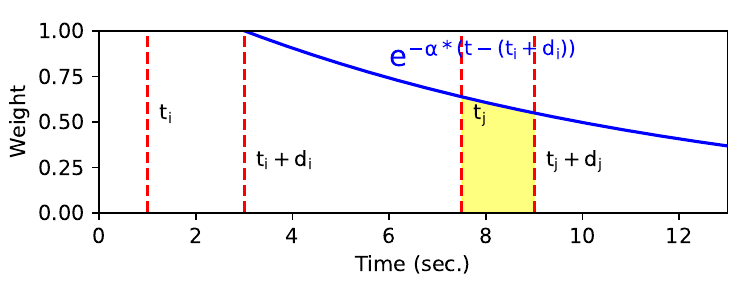}
  \caption{Example of two events where the yellow area corresponds to their contribution to the connection strength between from token $i$ to token $j$.}
  \label{fig:ex_integral_on_pair_events}
\end{figure}

\subsection{Ground Truth Interaction Matrix}
From each developer session, we derive a ground truth interaction matrix $\mS$.
For a fair comparison of neural models with developers, we take into account the tokenization used by the neural model, namely we use the CodeGen tokenizer~\footnote{\url{https://huggingface.co/docs/transformers/model_doc/codegen}} which is based on byte-level byte-pair-encoding~\cite{sennrichNeuralMachineTranslation2016}.

To convert char-level events into token-level ones, for each timestamp, if at least one character of a given token is visible, then the token is considered visible as well and we count the corresponding event $evt_{eye}^{(token)} = (t, i, d)$ where $t$ is the timestamp, $i$ is the token index and $d$ is the event duration.
Based on the pairs of events involving token $i$ and token $j$, we quantify how likely it is that the developer looks at token $j$ after having looked at token $i$.

Intuitively, we want to have stronger connection when a fixation on token $i$ is shortly followed by a fixation on token $j$, and if this second fixation has a significant duration.
Thus, we define the \textit{strength of the temporal connection} between token $i$ and token $j$ as:

\begin{equation}
   \emS_{i,j} = \sum_{evt_{i}, evt_{j} \in P_{i \rightarrow j}} \displaystyle \int_{t_j}^{t_j + d_j} e^{-\alpha (t - (t_i + d_i))} dt
\end{equation}
where $P_{i \rightarrow j}$ is the set with all the pairs of events where token $i$ is seen before token $j$ and the discounting factor $\alpha$ controls the decay of the connection the more the two events are far apart in time.
For our experiments we empirically set $\alpha = 0.1$, accounting for observed behavior where developers often spend several seconds scrolling and presumably shallowly searching the code.
In Figure~\ref{fig:ex_integral_on_pair_events} we show an example of the integral connecting two consecutive events.
\begin{figure}[t]
  \centering
  \includegraphics[width=.45\textwidth]{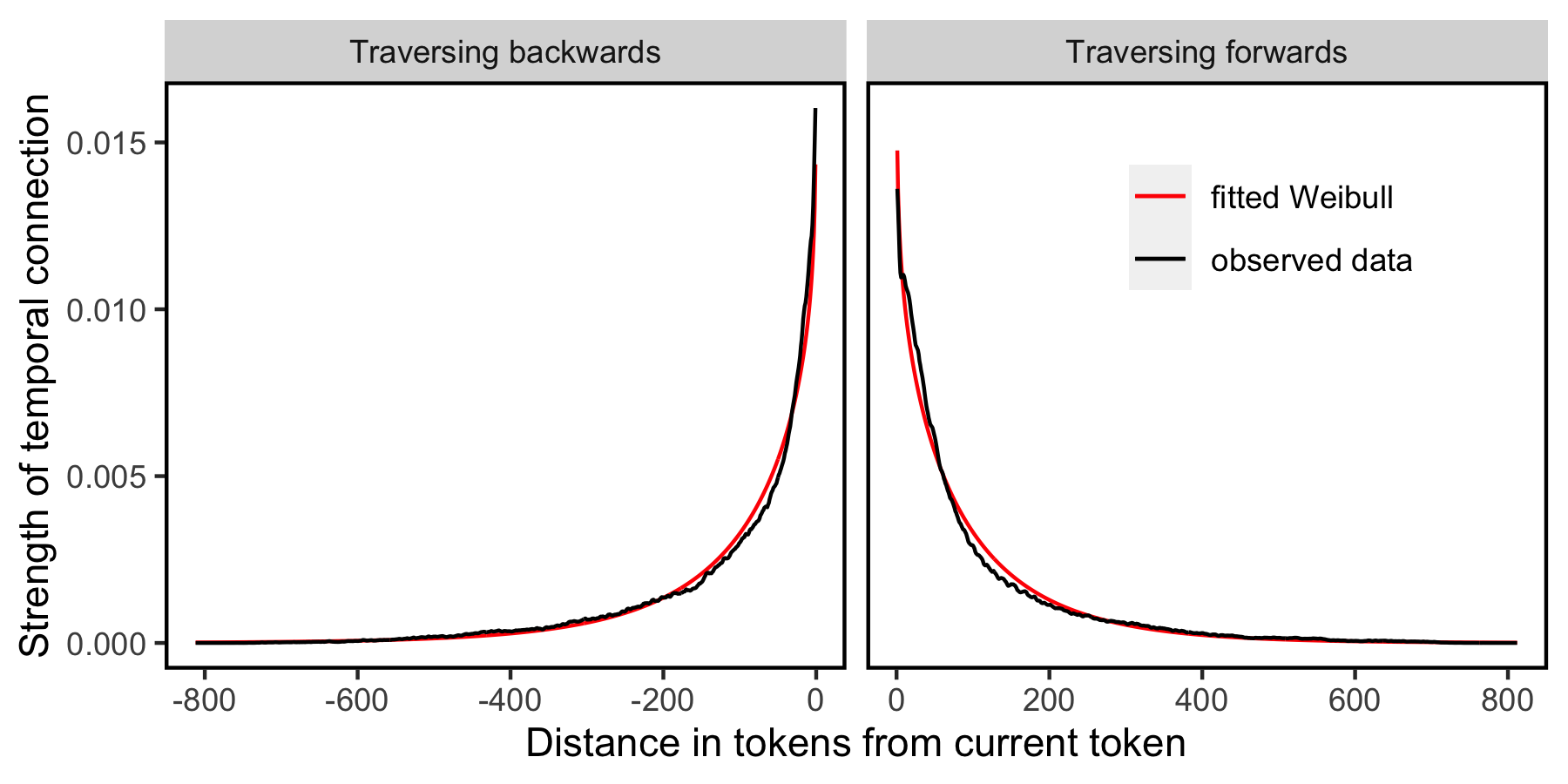}
  \caption{The strength of the connection $\emS_{i,j}$ depends significantly on the difference $i - j$. Both cases $i>j$ and $j>i$ can be well modelled using a Weibull distribution.}
  \label{fig:weibull}
\end{figure}

We noticed a strong \textit{neighboring effect} across the whole dataset, where the connection between closer tokens tends to be relatively stronger irrespective of context and content.
Indeed, developers do not jump randomly between likely locations within the code: they have a significant bias for staying close to their current position. We, therefore, begin by attempting to predict the observed strength of the temporal connection between tokens solely on the basis of their relative position.
We suggest a two-tiered approach: first, consider whether the developer is traversing forwards or backward, then use a relative model for \textit{how far} they will move in that direction.
We expect the ratio of forwards or backward traversal to be dependent on the exact task, and in fact, in our dataset the proportion of forward traversal ranged from \code{45.8\%} to \code{78.2\%}. For each task individually, as well as to some extent in general, the best simple predicting feature for traversal direction appears to be the current token position divided by the total number of tokens, i.e. the ratio of the document still in front of the developer. Fitting individual linear regressions for going forward (Eq.~\ref{eq:forward}) and backward (Eq.~\ref{eq:backward}) (which do not sum up to 1 because of the chance of returning to the token itself) gives the predictions of
\begin{align}
 \sum_{j > i}S_{i, j} \approx 0.94064 - 0.74599 * i / max(i) \ (R^2 = 0.56) \label{eq:forward}\\
 \sum_{j < i}S_{i, j} \approx 0.05261 + 0.74575 * i / max(i) \ (R^2 = 0.56) \label{eq:backward}
\end{align}
where the $R^2$ values indicate the goodness of fit, $i$ is the token index, $max(i)$ is the total number of tokens in the document, and the numeric values are the coefficients of the linear regression.
We expect, and find, the distribution for the distance between consecutive gaze points to be less dependent on the task. Of a number of standard distributions we tested against (normal, poisson, lognormal, exponential, Pareto, Weibull), it is best modeled using a fitted Weibull distribution, with the best fit of \code{shape = 0.89, scale = 98.14 tokens} going forward and \code{shape = 0.88, scale = 105.61 tokens} going backward (see Figure~\ref{fig:weibull}).

Thus, for the code exploration task, to extract relevance beyond mere closeness, we normalize each row of the interaction matrix $\mS$ by dividing by the average empirical ground truth distribution where the probability to go to a token constantly decreases the further away the target token is.

\section{Results}
\label{sec:results}

In this section, we compare the visual attention and interaction matrix extracted from the attention tensor of neural models against the ground truth computed from the developers.
We organize our empirical investigation in the following research questions:
\begin{itemize}
   \item \textbf{RQ1:} How effective are developers and neural models in solving sensemaking tasks?
   \item \textbf{RQ2:} How does the visual attention of developers and neural models compare?
   \item \textbf{RQ3:} How is the agreement between developers and neural models influenced by the programming language?
   \item \textbf{RQ4:} How do the interaction matrices of developers and neural models compare?
   \item \textbf{RQ5:} How is the effectiveness of \textit{follow-up attention} influenced by layer choice and number of newly generated tokens?
\end{itemize}
RQ1 and RQ2 considers three neural models: CodeGen~\footnote{https://huggingface.co/Salesforce/codegen-16B-multi}, GPT-J~\footnote{https://huggingface.co/EleutherAI/gpt-j-6B} and InCoder~\cite{friedInCoderGenerativeModel2022}.
Whereas, for the remaining questions we focus on the larger and more effective CodeGen model.

\subsection{RQ1: Answer Correctness}
\label{sec:rq1_answer_correctness}
\begin{figure}[t]
\centering
\includegraphics[width=0.47\textwidth]{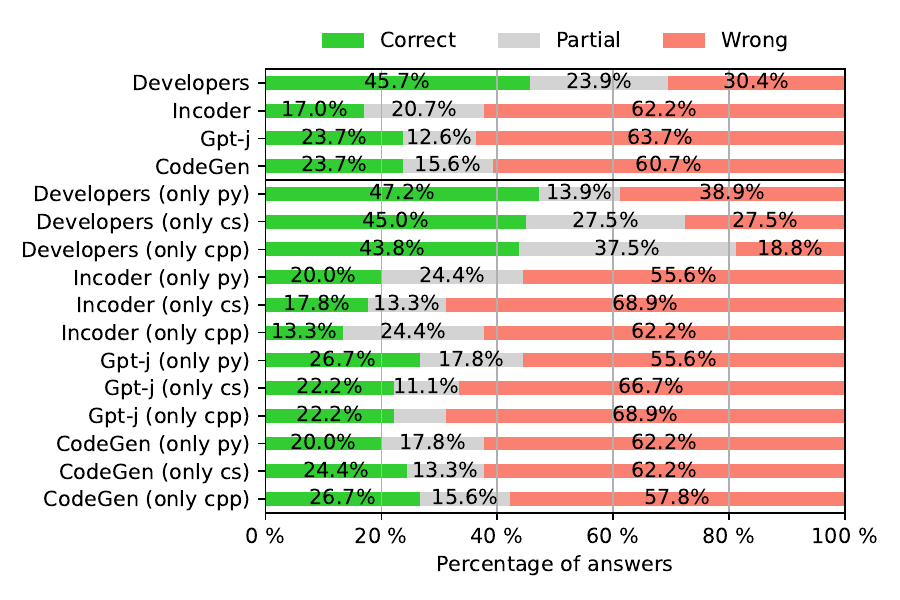}
\caption{Percentage of correct, wrong, and partially correct answers for developers and model.}
\label{fig:RQ1_RQ3_correctness}
\end{figure}

To evaluate the effectiveness of the developers and models in solving the sensemaking task, we annotate each generated answer by both groups involving four annotators in the process.
We use a scale of three values of correctness: (1) \textit{correct}, when the answer touches all the expected correct points, (2) \textit{partial}, if at least part of the correct answer is present or if the answer is wrong but in the same style of the correct solution (e.g. the Big-O notation), (3) \textit{wrong}, when the answer does not contain any correct part.
Note that, especially for the model, if the model generates extra text beyond the correct or partial answer we ignore the rest if it is incorrect.
Moreover, whenever the question is under-specified we accept multiple correct answers as long as they are compatible with the question.
To ensure a reproducible annotation process, all the authors collectively come up with a shared set of gold-standard answers for each question.
Then, two of the authors independently annotate more than 20\% of the answers generated by the model and the developer, and within two rounds of annotation followed by discussion, the final set of gold standard answers is agreed upon.
The final agreement on the 20\% of data led to a Cohen's Kappa of \cohenIRAHumanAnswers{}, \cohenIRACodeGenAnswers{}, \cohenIRAGPTJAnswers{} and
\cohenIRAInCoderAnswers{} for developers, CodeGen, Gpt-J, and InCoder respectively, which is considered a very high agreement~\cite{mchughInterraterReliabilityKappa2012}.
Finally, the remaining 80\% of the data is split in half and annotated only by one of the two authors individually.
The upper part of Figure~\ref{fig:RQ1_RQ3_correctness} shows the percentage of correct, partially correct, and wrong answers for each developer and neural model.

\begin{answerbox}
   \textbf{Answer to RQ1}: The developer's answers are correct or partially correct in almost 70\% of cases, while the neural models under study, namely CodeGen, Gpt-j, and InCoder, show a promising and non-trivial performance (respectively 39.3\%, 36.3\%, 37.7\% correct or partially correct answers).
 \end{answerbox}

\subsection{RQ2: Agreement on Visual Attention}
\label{sec:rq2_visual_attention}
\begin{figure}[t]
   \centering
   \includegraphics[width=0.45\textwidth]{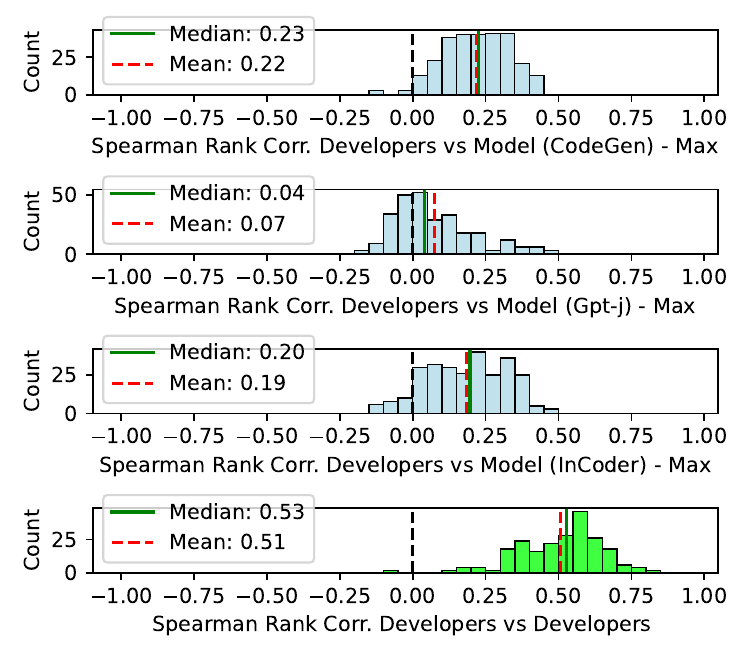}
   \caption{Agreement between developers and models' visual attention (extraction function: max).}
   \label{fig:human_vs_models_vector_agreement}
 \end{figure}

To measure the agreement between the visual attention of developers and the neural model, we regard them as vectors with meaningful ordinal content and compute their \textit{Spearman rank correlation coefficient}~\cite{spearmanProofMeasurementAssociation1987}, aligned with related work~\cite{paltenghiThinkingDeveloperComparing2021}.
In Figure~\ref{fig:human_vs_models_vector_agreement}, we report  the Spearman rank correlation between the developer attention vector and the model attention vector. For completeness, we also report the comparisons among developers.
Note that we only compare the attention maps of two different subjects from the different groups (e.g. developer vs CodeGen, developer vs GPT-J, etc.) when looking at the same code snippet and question.
Moreover, neither the data from the participants, nor those extracted by multiple model predictions are aggregates among subjects of the same group.
Instead, we consider the comparisons of all the possible combinations of subjects from the two groups.
We avoid aggregation because it may be sensitive to largely deviating data of single participants and the identification of a suitable aggregation function is a non-trivial task, as reported by~\cite{siegmundMasteringVariationHuman2021}. Related work~\cite{paltenghiThinkingDeveloperComparing2021} avoids aggregation for similar reasons.
This approach is adopted in all the comparisons among subjects in the paper.
The observed agreement exceeds that observed in previous work~\cite{paltenghiThinkingDeveloperComparing2021}, which we hypothesize to be due to a combination of us using a more advanced model~\cite{ahmadTransformerbasedApproachSource2020a} and a more natural data collection setup (eye tracking vs a deblurring interface).
To investigate whether higher model-human agreement is connected to higher effectiveness on the sensemaking task we compare agreement of the cases where both human and model are correct and where they are both wrong.
We run a statistical t-test and similarly to~\cite{paltenghiThinkingDeveloperComparing2021} we find that for InCoder the comparisons where both human and model are correct have a higher agreement than those where they both are incorrect (pval=2.23e-11).
We use the t-test under the assumption that the agreement values are normally distributed, which we confirmed by visual inspection.
For the other two models, there is no statistical significance.
\begin{answerbox}
   \textbf{Answer to RQ2}: The attention of neural models trained on code like CodeGen and InCoder exhibit a significantly higher agreement (+0.22, +0.20) with the developers when answering sensemaking questions as compared to GPT-j (+0.04), which was mainly trained on natural languages text.
 \end{answerbox}

\subsection{RQ3: Programming Language Analysis}
\label{sec:rq3_pl_analysis}

\begin{table}[t]
   \small
   \centering
   \begin{tabular}{lllrll}
      \toprule
        Model & Lang-A & Lang-B &   Stat. &     pval & Result \\
      \midrule
\rowcolor{RowGray}      Codegen &    cpp &     cs &  1486.0 & 9.95e-07 &  Diff. \\
         Gptj &    cpp &     cs &  3404.0 & 6.61e-02 &   Same \\
\rowcolor{RowGray}      Incoder &    cpp &     cs &  1280.0 & 1.96e-08 &  Diff. \\
      Codegen &    cpp &     py &  2685.0 & 7.22e-01 &   Same \\
\rowcolor{RowGray}         Gptj &    cpp &     py &  2571.0 & 9.37e-01 &   Same \\
      Incoder &    cpp &     py &  2947.0 & 1.73e-01 &   Same \\
\rowcolor{RowGray}      Codegen &     cs &     py & 10334.0 & 9.29e-15 &  Diff. \\
         Gptj &     cs &     py &  5323.0 & 2.00e-02 &  Diff. \\
\rowcolor{RowGray}      Incoder &     cs &     py & 10327.0 & 1.04e-14 &  Diff. \\
      \bottomrule
      \end{tabular}

   \caption{Results of the Mann-Whitney U statistical tests when comparing the distributions of developer-model agreement (Spearman rank coefficients) across models.}
   \label{tab:pl_diff_statistical_test_results_all_models}
\end{table}

We investigate the differences in the agreement between developers and models across programming languages.
In the lower part of Figure~\ref{fig:RQ1_RQ3_correctness} we report the answer correctness (RQ1) divided into groups across the three programming languages under study: C\#, C++, and Python.
Since each developer has participated in the study only using a single programming language, the difference in answer correctness of developers might  be a result of both the programming language or the skill and background of the specific developer.
On the other side, the neural models are equally applied to the different languages, and thus we expect that the difference in answer correctness is due to the specific programming language.

We compare the agreement between developers and models across programming languages with the Mann-Whitney U test~\cite{mannTestWhetherOne1947} to compare two distributions.
Table~\ref{tab:pl_diff_statistical_test_results_all_models} illustrates that the agreement between developers and models is significantly higher on C\# than on Python (p-value $< 0.05$), and marginally significantly higher on C\# than on C++ (p-value $< 0.1$), making C\# the language with the highest agreement between developers and models.

\begin{answerbox}
   \textbf{Answer to RQ3}:
   The effectiveness in answering sensemaking questions is influenced by the specific programming language in which the task is formulated to the neural model, with a gap of up to 13.3 absolute points.
   Whereas, the agreement between developers' attention and neural models is higher for C\# than for Python, and marginally higher for C\# than for C++.
 \end{answerbox}

\subsection{RQ4: Agreement on Interaction Matrix}
\label{sec:rq4_interaction_matrix}

\begin{figure}[t]
   \centering
   \includegraphics[width=0.45\textwidth]{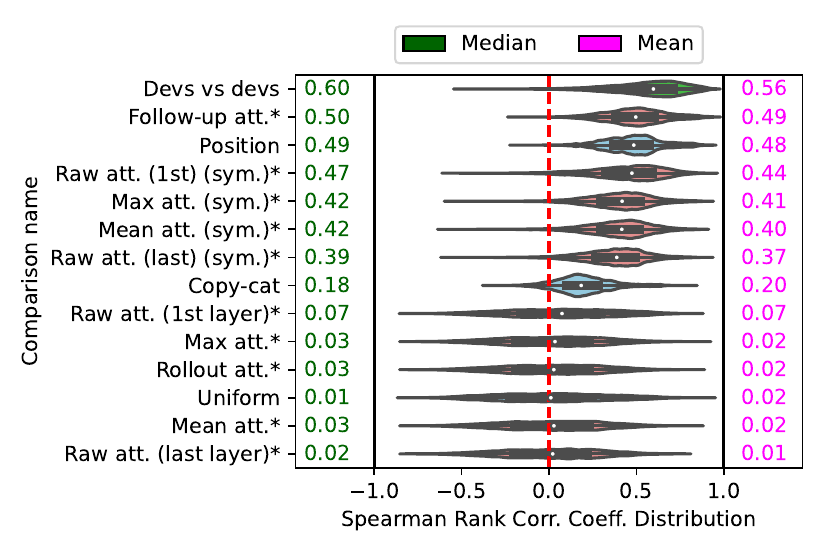}
   \caption{Spearman Rank Correlation Developers vs Codegen interaction matrix (* for attention-based method) .}
   \label{fig:tl_spearman_all_comparisons}
\end{figure}

Considering CodeGen's better answer effectiveness and higher developer agreement, we restrict our subsequent investigation to this model.

To compute the agreement between the interaction matrices we use a row-by-row approach, thus considering the task from the perspective of the starting location and asking ourselves: \textit{to which code location should I look next, given that now I am looking at this token?}
As a possible target code location, we consider the other lines in the code, thus we compare line-level interaction matrices $\mS_{line}$. We focus on lines for two reasons: a) because it is well-known that LLMs attend to different kinds of tokens within a line than humans, such as punctuation or newlines\cite{paltenghiThinkingDeveloperComparing2021}, and b) because we argue that the line-level is perhaps the most useful granularity from a hypothetical user perspective (see Section \ref{sec:conclusion}).
To obtain $\mS_{line}$, we take $\mS$ and sum the probabilities referring to tokens on the same line, thus the probability to go to a line $x$ is the probability to go to any token of that line.
To compare corresponding rows of the ground truth interaction matrix and that derived by our attention signal, we use both the Spearman rank correlation coefficient or the top-3 overlap, defined as the number of top-3 target positions shared between the ground truth row and the model-derived row.
Moreover, to balance the fact that some potential starting tokens might be rarely (or only very transiently) looked at by the developers, we weight each comparison based on the total number of seconds spent by the developer on the corresponding starting token.
We also fix a maximum for this weight to 10 sec to prevent long-observed tokens from dominating the comparison.

We run the different extraction functions on the model attention signal to obtain interaction matrices $\mS_{line}$, which we compare to developer-derived ground truth.
We distinguish between: (1) \textit{attention-based code traversal} predictions, which are those introduced in Section~\ref{sec:extract_functions_interaction_matrix}, and (2) \textit{attention-agnostic code traversal} predictions.
The attention-based methods comprise raw attention in the first and last layer, max, and mean, with their respective symmetric versions where the triangular matrix is mirrored and added to replace the zero values, the rollout, and follow-up attention.
The attention-agnostic methods comprise: \textit{copycat} recommending all the positions containing tokens identical to the starting token (e.g., starting from token \code{print} it recommends all other lines containing \code{print} with equal weight), \textit{uniform} recommending all the positions preceding the current token, and \textit{position} recommending the neighboring positions of the current token with a Gaussian distribution centered on the current token.

\begin{figure}[t]
   \centering
   \includegraphics[width=0.45\textwidth]{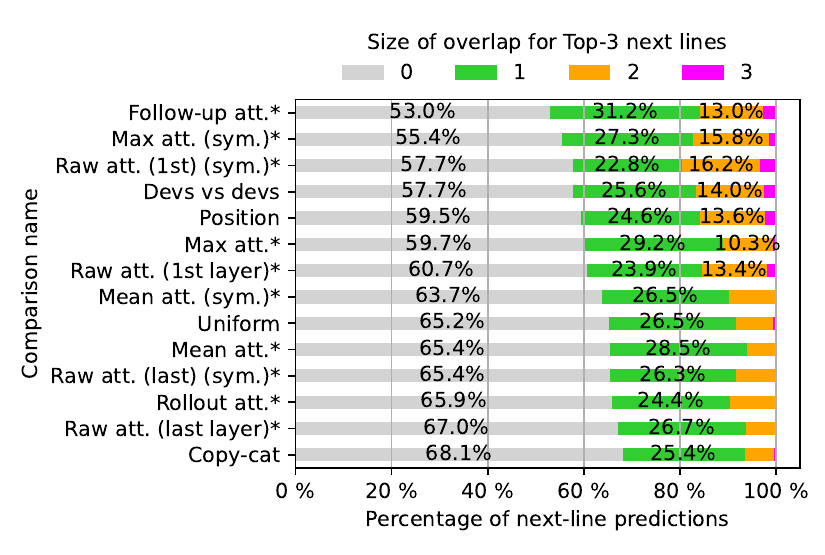}
   \caption{Top-3 Overlap Developers vs Codegen interaction matrix (* for attention-based method).}
   \label{fig:tl_top3_all_comparisons}
\end{figure}

We find that attention-based methods do carry predictive power, and in particular that \textit{follow-up attention performs best} among all methods for both Spearman rank and top-3 overlap (Figures~\ref{fig:tl_spearman_all_comparisons} and \ref{fig:tl_top3_all_comparisons}).
We note that a purely position-based approach performs better than the copycat method despite being completely content-agnostic.
We attribute this to developers' tendency to often read source code in (piecewise) linear order as described by \cite{blascheckVisuallyAnalyzingEye2019}.
Regarding raw attention, \cite{zhangWhatDoesTransformer2022} demonstrated deeper semantic information being concentrated in later layers. Yet for both the triangular and symmetric versions respectively, higher levels appear inferior at predicting eye movement to earlier levels, possibly because such deeper semantic information may not be apparent to developers.

\begin{answerbox}
   \textbf{Answer to RQ4}:
   The follow-up attention function performs best in predicting the next code location to look at, with a Spearman rank correlation of +0.49 and a top-3 overlap of 47\%. This outperforms the baseline prediction
   accuracy of 42.3\%, which uses the session history of another
   developer to recommend the next line.
\end{answerbox}

\subsection{RQ5: Ablation Study}

\begin{figure}[t]
   \centering
   \includegraphics[width=0.45\textwidth]{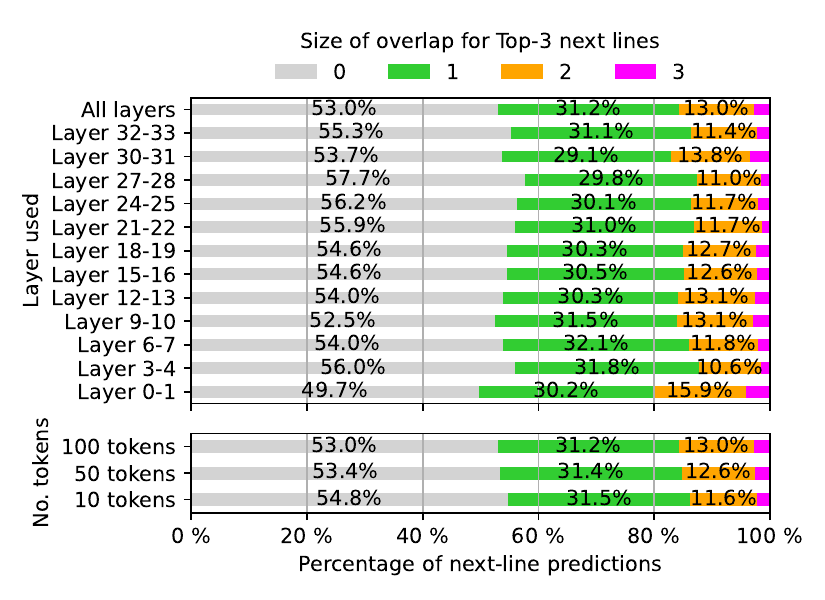}
   \caption{Effect of layer pair and number of generated tokens on top-3 overlap.}
   \label{fig:tl_top3_abl_layers_and_tokens}
\end{figure}

\label{sec:ablation}
We investigate two key design choices for follow-up attention: (1) the selection of layers to use, and (2) the number of generated tokens, i.e., observers.
In the top part of Figure~\ref{fig:tl_top3_abl_layers_and_tokens}, we report the top-3 overlap for different pairs of layers among the 34 available in CodeGen, together with the configuration where all layers are considered on top.
In the same lower part of Figure~\ref{fig:tl_top3_abl_layers_and_tokens}, we show the top-3 overlap when restricting the usage of the next 10, 50, or 100 generated tokens or ``followers''.
There is significant agreement with the ground truth even for smaller numbers of layers, particularly the very first one. This suggests little processing, and maybe even only the token embedding, might be needed to extract valuable information. Additionally, a higher number of observers of the follow-up has a positive impact on the agreement of the follow-up attention with the ground truth.

\begin{answerbox}
   \textbf{Answer to RQ5}:
   The follow-up attention benefits more from using the attention signal produced by early layers and performance are robust to the number of generated tokens.
\end{answerbox}

\section{Threats to Validity}
\label{sec:threats_to_validity}

There are potential threats to validity that may limit the generalizability of our findings.
First, our sample of developers may not represent all such developers, especially since we recruited from one large technology company. However, we did screen participants and require that they have professional programming experience. Second, the tasks are sensemaking tasks, as opposed to ecologically valid debugging or feature enhancement tasks, involving code that participants are not familiar with, and the participants were restricted from running the code, using a debugger, and performing web searches. Such tasks are commonly used in technical interviews and the participants did not indicate the tasks were atypical.
Third, reactivity effects may occur since participants knew they were being observed and may believe their technical ability is being assessed. To minimize this threat, we advised participants that their individual performance was not being reported or analyzed and the observations were performed remotely with the researcher not being present in the same physical room as the participant. Fourth, it is possible that a developer's gaze does not always represent their attention, though such eye-tracking data has been well-studied for decades in numerous domains.
Fifth, regarding possible model's ``cheating'' due to memorization~\cite{yangUnveilingMemorizationCode2024}, we acknowledge the impossibility to exclude that these programs have not been seen by the models during training.
However, we note that we evaluate them on the performance on the sensemaking task, for which the combination snippet and question is novel to the models since it was created for this study.
Regarding the attention distribution, whether models exhibit different attention patterns on code snippet that have been during training, as opposed to those that are novel to them, still remains an open question for future work.
Finally, the answers generated by the neural models are dependent on the prompts we provide, and thus results may vary with more elaborate prompt design, which is an active research area in prompt engineering~\cite{prennerAutomaticProgramRepair2021,
abukhalafCodexPromptEngineering2023, caoStudyPromptDesign2023,shrivastavaRepositoryLevelPromptGeneration2022b,reynoldsPromptProgrammingLarge2021,bareissCodeGenerationTools2022}.

\textbf{Generalizability}.
In the design of the study, we aimed to make the evaluation as generalizable as reasonably possible and in line with the current state of the art in the field.
On the human side, comparing our sample size to what is found in other eye-tracking studies~\cite{obaidellahSurveyUsageEyeTracking2018}, we note that both number of participants (25 participants vs $\mu=19.6$, $\sigma=13.5$) and number of programs (15 programs vs $\mu=7.6$ programs, $\sigma=17.2$) are in line with other studies.
On the contrary, most of the current literature focuses on Java (38\%)~\cite{obaidellahSurveyUsageEyeTracking2018}, while we have a more diverse set of programming languages (C\#, C++, and Python), possibly making our results more generalizable.
On the model side, we picked a diverse set of widely popular models in terms of downloads on HuggingFace: CodeGen (81K+), GPT-J (2.5M+), and InCoder (58K+).
Regarding the applicability of the follow-up attention to other models, especially closed-source ones, although current model inference APIs do not expose attention information yet, we note that the vast majority of closed source are transformer-based making the extraction of attention possible in principle.

\section{Discussion and Implications}
\label{sec:discussion}

Generally, each part of a codebase holds myriad disparate connections to other parts of the codebase in such forms as documentation, calls and tests, pattern and format parallelism, examples, data, and control flow.
However, the positive moderate Spearman Rank correlation among developers (+0.56) shows that the developers tend to navigate a single file along similar paths when trying to make sense of it with the same goal, i.e. answering to the same question.
To some extent, this points to a common notion representing a relationship of general \textit{``relevance''} of one location to the other, at least as far as we consider a single file as done in the current study; more work is needed to generalize to larger codebases and across files.
At the same time, we also find this general relevance relation in human understanding is reflected in the neural processing of large transformer models of code, which also show a remarkably promising correlation with the human exploration paths (+0.49).
Surprisingly, the follow-up attention agrees with the ground truth on what line to look at next even more than the developers agree with each other: 47\% vs 42.3\% (Figure~\ref{fig:tl_top3_all_comparisons}).
Note that the two runner-up are still attention based and they also outperform or match developers' agreement: 44.6\% for \textit{Max att. (Sym)} and 42.3\% for \textit{Raw att. (1st) (Sym)}.
This shows that attention-based approaches, and follow-up attention above all, are promising for recommending the next code location to look at.
Thus, this motivates further work on the analysis and use of the neural attention layers as promising way to support developers in their code exploration tasks.

\textbf{IDEs with 360$^{\circ}$ Vision.}
Developers spend a large part of their time understanding existing code~\cite{Ko2006TSE, Piorkowski2016FSE}, and a central role of advanced code editing environments is to facilitate navigation to relevant places, whether the developers' involvement is active (e.g., search), passive (e.g., highlighting of identical tokens) or semi-active (e.g., jump-to-definition).
Such tools typically rely on a proxy for current developer focus, such as mouse pointer position or the user's cursor in a code editor.
The challenge is to find the locations relevant to that focus location since the possible reasons for relevance are heterogeneous and syntactical methods can only surface a limited number of them.
Nevertheless, such tools have been an active research area~\cite{Robillard2010Software}, with results such as Strathcona~\cite{Holmes2005ICSE}, Suade~\cite{Warr2007ICSE}, Team Tracks~\cite{DeLine2005VLHCC}, Navtracks~\cite{Singer2005ICSM}, Mylar~\cite{Kersten2005AOSD}, PFIS~\cite{Piorkowski2012CHI}, Prodet~\cite{Augustine2015ICSE}, and Hipikat~\cite{Cubranic2003ICSE}.
In fact, Singh et al. evaluated various operationalizations of human attention (i.e., cursor location, which code is visible on screen, and a qualitative human judgment) and its impact on predictive accuracy~\cite{Singh2016ICSME}, though they did not include eye tracking data.
The high rate of success of follow-up attention of recommending at least one relevant line among the top-3 (47\%) shows the effectiveness of the attention of neural in providing one such possible proxy for relevant code locations connected to the current statement.
Further research is needed to explore how to best incorporate such a proxy into existing tools, and how to best use it to support developers in their code exploration tasks, e.g. either highlight neural attended lines, offer to link to them, or list them in a side panel.

\textbf{LLMs and Human Collaboration.}
Although the sensemaking task spans over diverse set of topics and has mostly open-ended questions, CodeGen, the largest LLM studied achieves already non-trivial performance with 39.3\% of correct or partially correct answers.
This is a promising result for the use of LLMs in supporting developers when reasoning on code, and motivates further research on perhaps more specialized sensemaking questions directly liked to specific traditional software engineering task, such as ``Is there a bug in this code?'' for bug detection~\cite{pradelDeepBugsLearningApproach2018, habibNeuralBugFinding2019} or ``Is this code vulnerable to SQL injection?'' for vulnerability detection~\cite{chakrabortyDeepLearningBased2022}.

\textbf{Context Prioritization for LLMs.}
In existing tooling employing LLMs, such as GitHub Copilot~\cite{GitHubCopilotYour}, the model can process only limited part of the code at the same time, given by the maximum size of the prompt it can take, also called \emph{context window}~\cite{EddieAlbertDagstuhl, shrivastavaRepositoryLevelPromptGeneration2022b}.
In practice, heuristics are needed to explore which parts should be included in the prompt.
Our findings show how attention-based methods exhibit a moderate positive agreeement with human experts, especially on the top-3 next lines (47\%), thus could be used to prioritize the context to be included in the prompt.
From our ablation study, it is encouraging to see that even with only 10 newly generated tokens, the agreement is still higher than the agreement between developers: 45.2\% vs 42.3\%.
These results suggest that real-life deployment of the follow-up attention as a relevance provider could benefit from two important optimizations to reduce the computational cost: (1) restricting the number of layers to consider to the first two, and (2) restricting the number of generated tokens to consider.

\section{Conclusion}
\label{sec:conclusion}
We presented and shared a novel dataset of eye-tracking data, comprising \nEyeTrackingSessions{} visual attention sessions of \nEyeTrackingParticipants{} developers when answering sensemaking questions in three popular programming languages (Python, C++, and C\#).
We confirmed that neural models provide promising but less accurate answers than developers to these questions while paying attention to similar parts of the code.
We formalized a new code exploration task of predicting developer code traversal and confirmed the attention signal's relevance for this task by evaluating multiple processing approaches.
Besides evaluating existing approaches on the sensemaking task, we contributed the concept of follow-up attention, which shows the best agreement with the developer attention data.

\section{Data Availability}
\label{sec:data_availability}
All our code is publicly available at \url{https://github.com/githubnext/followup-attention} and the dataset is available here~\footnote{Raw eye tracking sessions \url{https://doi.org/10.6084/m9.figshare.23599251} and clean version \url{https://doi.org/10.6084/m9.figshare.23599233}}

\section{Acknowledgments}
We thanks our colleagues at GitHub and Microsoft for their support and feedback on this work. We also thank the anonymous reviewers for their valuable feedback and suggestions.

\balance
\bibliographystyle{IEEEtran}
\bibliography{phd-mattepalte, iclr2023_conference}

\begin{IEEEbiography}[{\includegraphics[width=1in,height=1.25in,clip,keepaspectratio]{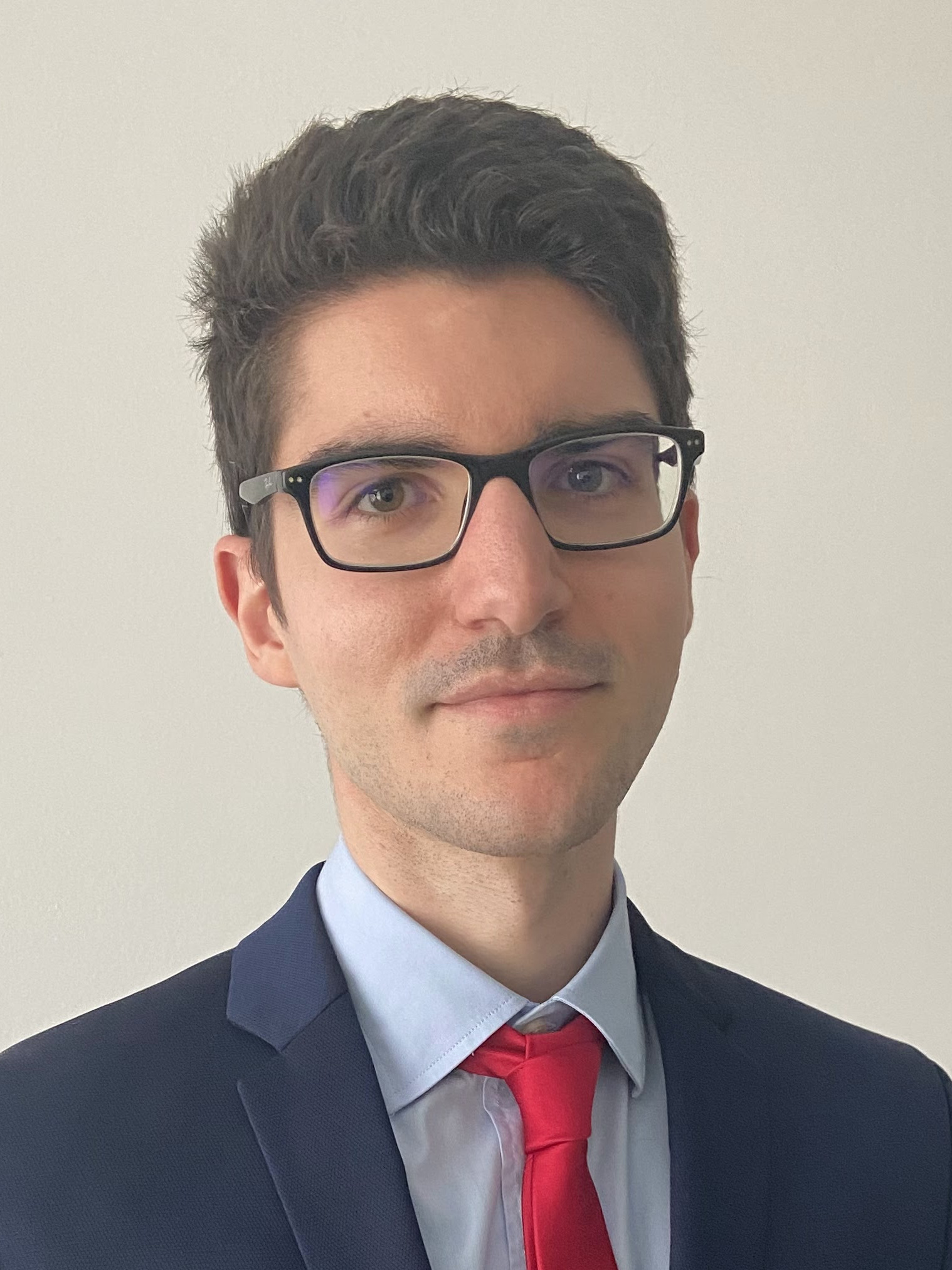}}]{Matteo Paltenghi}
is a doctoral researcher at the University of Stuttgart with expertise at the intersection of artificial intelligence and software engineering, advised by Prof. Dr. sc. Michael Pradel. Beside his Ph.D., he also worked as research scientist at GitHub Next and CERN. He is an active member of the software engineering community where he serves as reviewer (TSE, TOSEM, JSSoftware), and his work was also presented at top conferences like ASE 21, OOPSLA 22, ICSE 23 and ICSE 24.
Matteo holds a double degree M.Sc. in Computer Science and Engineering from Politecnico di Milano and TU Berlin. Recently, he was selected as young researchers for participation in the Heidelberg Laureate Forum (HLF 23). For more information, visit \url{https://matteopaltenghi.com}.\end{IEEEbiography}

\begin{IEEEbiography}[{\includegraphics[width=1in,height=1.25in,clip,keepaspectratio]{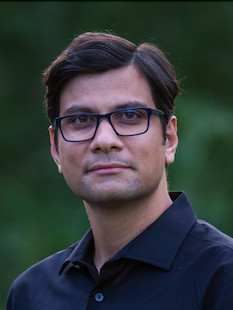}}]{Rahul Pandita}
is staff researcher at GitHub Inc. where he works on automated software engineering at with focus on devloper tools and developer productivity. He was previously a Senior Researcher at Phase Change Software where he developed a virual assistant `MIA' to help COBOL developers comprehend legacy mainframe systems. He received his Ph.D. of Computer Science from North Carolina State University in 2015. For more information visit \url{http://rahulpandita.me/}\end{IEEEbiography}

\begin{IEEEbiography}[{\includegraphics[width=1in,height=1.25in,clip,keepaspectratio]{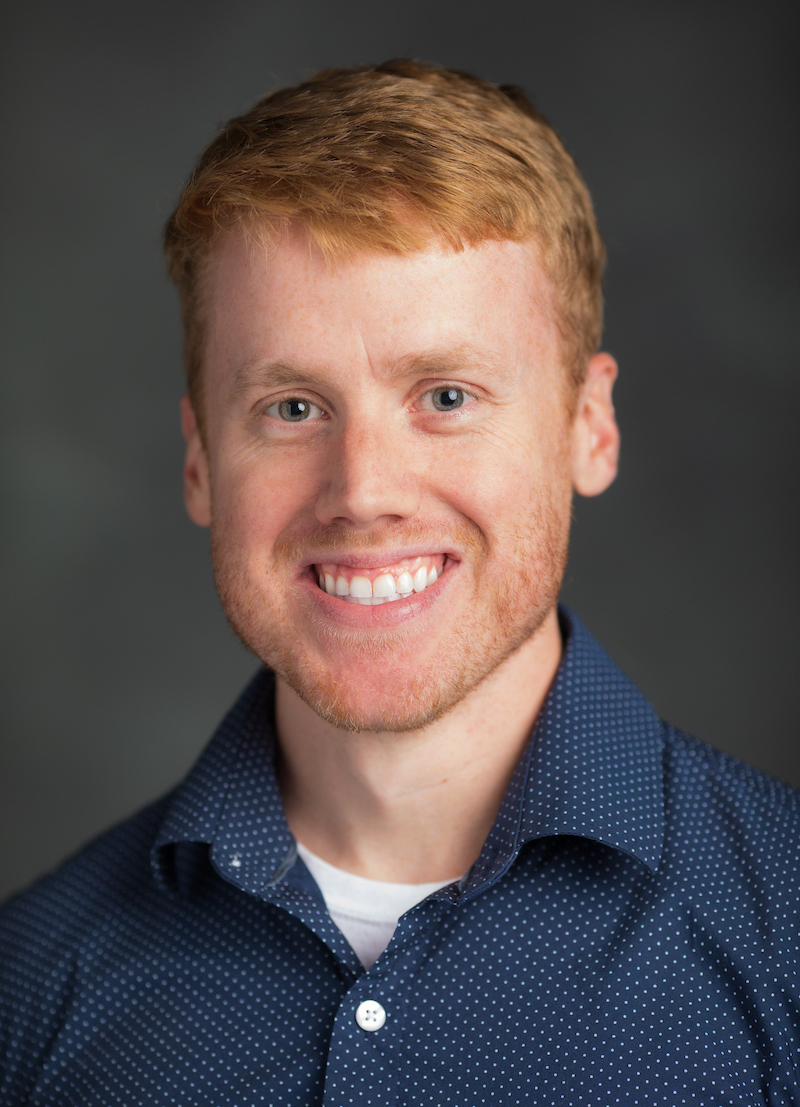}}]{Austin Z. Henley}
is a Senior Researcher at Microsoft where he works on the human factors of AI-powered developer tools. Previously, he was a tenure-track professor at the University of Tennessee where he led an NSF-funded lab researching developer productivity and taught software engineering courses. He received his Ph.D. in Computer Science from the University of Memphis in 2018. For more information, visit \url{http://austinhenley.com/}.\end{IEEEbiography}

\begin{IEEEbiography}[{\includegraphics[width=1in,height=1.25in,clip,keepaspectratio]{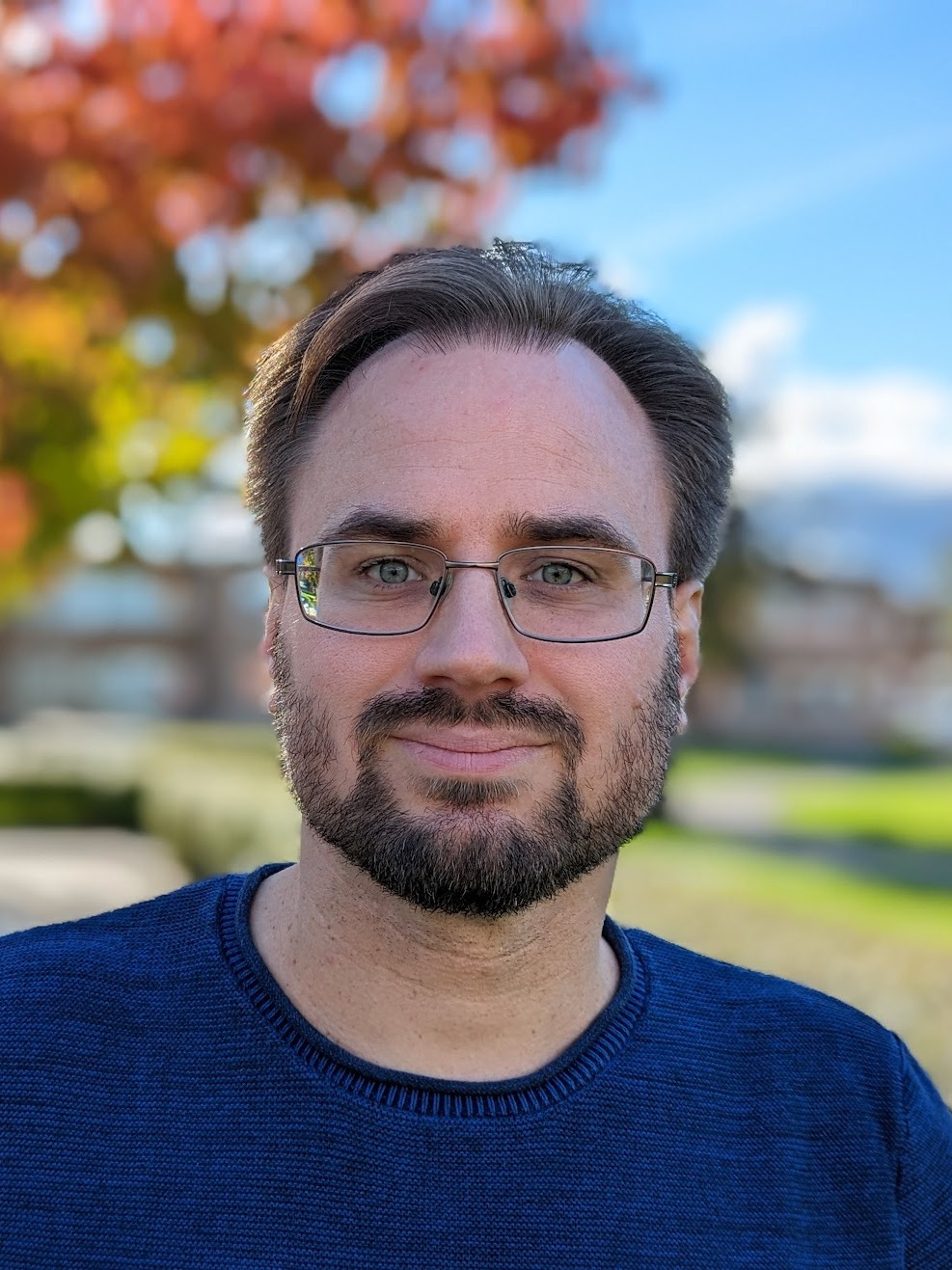}}]{Albert Ziegler}
is a Principal Researcher at GitHub Next
where he works on Artificial Intelligence for the Software Development Lifecycle.
He was one of the three original inventors of GitHub Copilot and has since
turned to LLM guided tooling in both in the IDE (Copilot NES, Copilot Radar)
and the pull request workflow (AI for Pull Requests, Gentest).

He holds a PhD in Mathematics from Leeds University
and has previously worked on developer productivity
and diverse industry ML projects.\end{IEEEbiography}

\end{document}